\documentclass[aps,pra,amsmath,amssymb,twocolumn,groupedaddress,showpacs,showkeys]{revtex4-1}

\usepackage{latexsym}
\usepackage{amssymb}
\usepackage{amsmath}
\usepackage{graphicx}
\usepackage{dcolumn}
\usepackage{bm}

\begin{document}


\title{Large-momentum-transfer Bragg interferometer with strontium atoms}



\author{T. Mazzoni}
\author{X. Zhang}
\email{also: International Centre for Theoretical Physics (ICPT), Trieste,Italy}
\author{R. Del Aguila}
\author{L. Salvi}
\author{N. Poli}
\author{G. M. Tino}
\email{Guglielmo.Tino@fi.infn.it}
\affiliation{Dipartimento di Fisica e Astronomia and LENS -
Universit\`{a} di Firenze, INFN - Sezione di Firenze, Via Sansone
1, 50019 Sesto Fiorentino, Italy}





\date{\today}

\begin{abstract}
We report on the first atom interferometer based on  Bragg diffraction in a fountain of alkaline-earth atoms, namely \textsuperscript{88}Sr. We demonstrate large momentum transfer to the atoms up to eight photon recoils and the use of the interferometer as a gravimeter with a sensitivity  $\delta g/g=4\times
10^{-8}$.
Thanks to the special characteristics of strontium atoms for precision measurements, this result opens a new way for experiments in fundamental and applied physics.
\end{abstract}

\maketitle


\section{Introduction and motivation}

Atom interferometers are rapidly evolving, being used as new quantum sensors
for fundamental physics experiments and in several other applications
\cite{Varenna2013}. In gravitational physics, for example, they
enable precise measurements of gravity \cite{Peters1999,Gillot2014}, gravity
gradients \cite{McGuirk2002,Sorrentino2014}, gravity curvature
\cite{Rosi2015}, and of the Newtonian gravitational constant
\cite{Rosi2014}.
Important goals are the
increase of their sensitivity and the demonstration of interferometry with
atomic species other than alkali atoms, which are most commonly used. For some experiments, indeed, the possibility of choosing the atomic species with the right characteristics is crucial. In particular, for precision measurements there is a considerable interest in using
alkaline-earth or alkaline-earth-like atoms, such as Ca, Sr or Yb \cite{Riehle1991,Tarallo2011,Graham2013,Tarallo2014,Jamison2014,Hartwig2015}, that are already used for the 
most advanced optical atomic clocks \cite{Hinkley2013,Ushijima2015,Bloom2014,Poli2013}.
Alkaline-earth atoms have  several characteristics that make them particularly interesting in this context.
Firstly, their zero electronic angular momentum
in the $^1$S$_0$ ground state makes these atoms less sensitive to perturbation due to magnetic fields than alkali atoms.
Furthermore,  they offer more flexibility  thanks to the presence of both
dipole allowed transitions and narrow intercombination
transitions that can be used for efficient multiphoton Bragg diffraction
\cite{Giltner1995,Muller2008b,Altin2013} and for single-photon atom
interferometry schemes \cite{Yu2011,Graham2013}. 
Finally, resonance transitions from the ground state are in the blue/near-UV  (e.g., 461 nm for
Sr, 399 nm for Yb)  resulting in a larger momentum transferred to the
atoms for the same diffraction
order compared to alkali atoms  and hence in a correspondingly higher potential sensitivity of the interferometers.


Here, we demonstrate the first atom interferometer based on large
momentum transfer (LMT) Bragg diffraction in a fountain of alkaline-earth
atoms, namely strontium, and its use for the measurement of gravity acceleration. 
%
\begin{figure}[b]\begin{center}
\includegraphics[width=0.47 \textwidth]{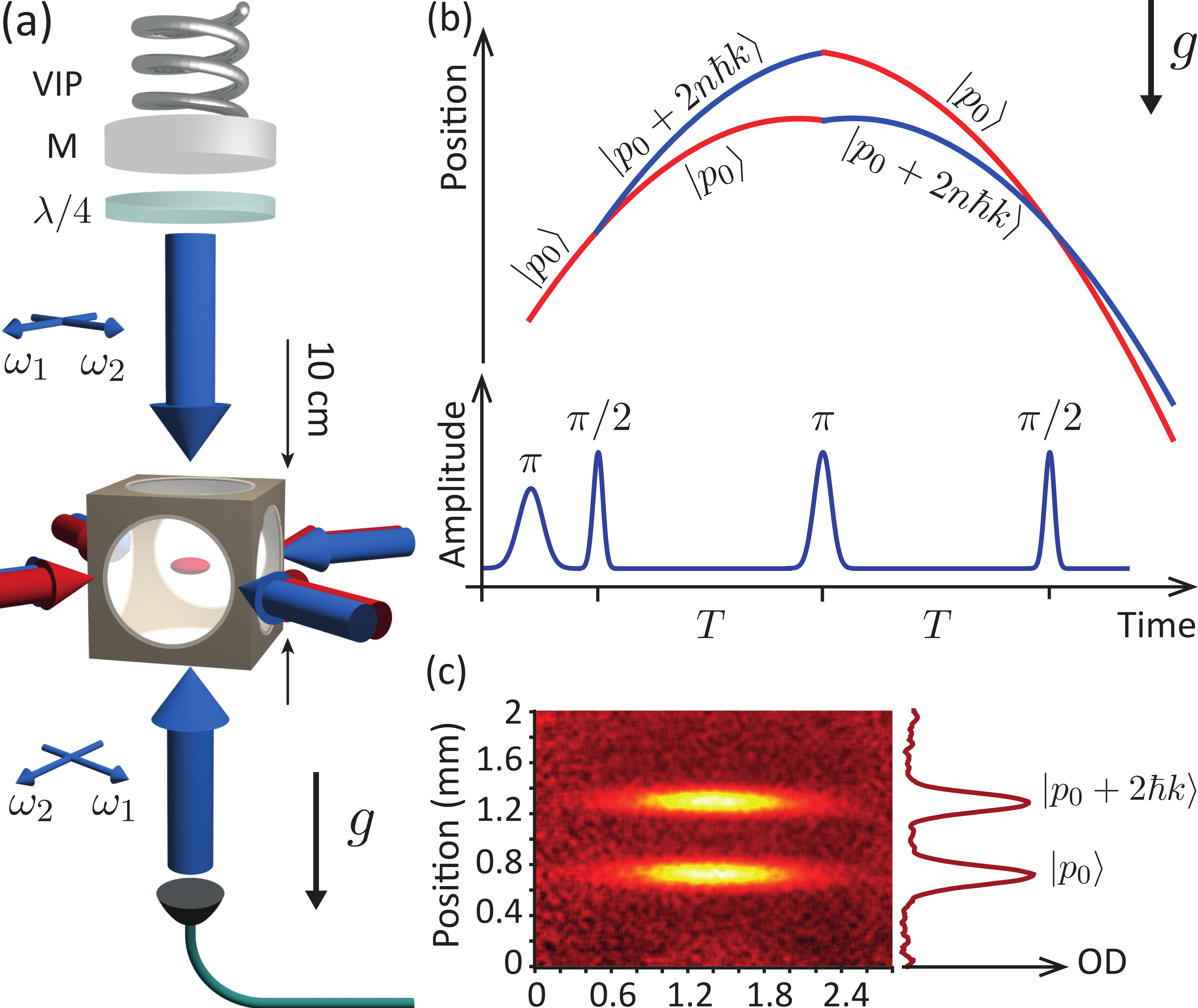}
\caption{(a) Simplified picture of the experimental apparatus. The \textsuperscript{88}Sr atoms
are cooled in a double-stage magneto-optical trap. The Bragg laser beams  with frequencies $\omega_1$
and $\omega_2$ and orthogonal polarizations are sent vertically from the bottom
of the chamber, rotated by a $\lambda/4$ wave-plate and
retro-reflected by a mirror (M) installed on a
vibration isolation platform (VIP). (b) Scheme of the atom interferometer with separated arms corresponding to different momentum states under the effect of gravity.
Before the interferometric sequence the atoms are
velocity selected and launched by a sequence of $\pi$ pulses.
(c) Time of flight image ($T_{\textrm{tof}}=30$ ms) of the two interferometer arms split
by a 1\textsuperscript{st} order $\pi/2$ pulse. The spatial separation
after 30 ms  is 600~$\mu$m.} \label{fig.apparatus}
\end{center}
\end{figure}
In addition to the general features of alkaline-earth atoms listed above, the \textsuperscript{88}Sr isotope that we use in this work has specific favorable characteristics: it has no nuclear spin so that in the ground state it is a scalar particle which is virtually insensitive to stray  magnetic
fields, 
and its small scattering length $a=-2a_0$ \cite{Ferrari2006,Martinez2008,Stein2010} results in reduced decoherence due to cold collisions.
This allows, for example, observation of extremely long-lived Bloch oscillations  of \textsuperscript{88}Sr atoms in a vertical optical lattice \cite{Ferrari2006,Poli2011}.
On the other hand, since strontium has no hyperfine structure in the ground state,
the usual schemes based on Raman transitions cannot be employed to realize the
 beam splitters and the mirrors for an interferometer. 
 In this work, we use Bragg diffraction which acting only on the atom's external degrees of freedom can  split the atomic wavepacket into two momentum states separated by
$2n\hbar k$ (where $n$ is the Bragg diffraction order, and $k=
2\pi/\lambda$ is the wavevector of the Bragg laser light with a wavelength $\lambda$), while maintaining the same electronic state.  

\section{Method and experimental setup}

In Fig.\ref{fig.apparatus}(a), a schematic view of the
experimental apparatus is shown. 
The beams at 461 nm for the Bragg transitions are produced by a home-made  laser that is frequency-locked to the main cooling laser with a red detuning $\Delta$ which is set, for
different Bragg orders, in the  $3-8$ GHz range with respect to the 
$^{1}S_{0}$--$^{1}P_{1}$ transition frequency. The output power is
about 200 mW and the emission linewidth is about 1 MHz. The 
laser intensity is actively stabilized using an external single-pass
acousto-optical-modulator (AOM) 
(see appendix for details on the noise spectrum).
The two Bragg beams, with frequencies $\omega_1$ and $\omega_2$,
are obtained using two separate AOMs and they are coupled with mutually orthogonal polarizations into a single-mode polarization-maintaining
fiber. They are
collimated at an $1/\mathrm{e}^2$ intensity radius of $r=2.5$ mm
and sent vertically upwards onto the atomic sample. The light is then retro-reflected by a $2"$ mirror
suspended on a vibration isolation platform (MinusK 25BM-4). A
quarter-wave plate is  placed before the retro-reflection
mirror to rotate the polarization of the returning light by
$90^{\circ}$. This allows the beams to interfere with each other
to generate two travelling waves moving in opposite directions,
while the formation of standing waves by pairs of beams with the same
frequency is avoided. The difference between the beams' frequencies
$\delta_n=\omega_1-\omega_2$ is adjusted in order to have the upward moving lattice drive the Bragg
transitions,
which occur for $\delta_n= 4n \omega_r$  in the falling frame, where $\omega_r=\hbar k^2/2m = 2\pi \times
10.7$ kHz is the recoil frequency for strontium atoms. The  lattice  moving  downward is Doppler shifted
out of resonance during most of the atoms' free-fall. Bragg pulses at the apogee
of the ballistic trajectories are  avoided to prevent
double diffraction. The verticality of the beam is verified at
1~mrad  by retro-reflecting it on a water surface.
The residual vibrations and tilt coupled to the retro-reflecting mirror
are monitored by a triaxial accelerometer (Episensor
ES-T) and a precision tiltmeter (Applied Geomechanics Tuff Tilt
420) placed on top of the vibration-isolation platform
(see appendix for details on noise spectra).
The whole platform is enclosed in an acoustic
isolation box.
The two Bragg AOMs are driven by two radio-frequency (RF) generators
phase-locked to a 10 MHz reference signal provided by a Rb clock and the
pulses are shaped to have a Gaussian profile \cite{Muller2008}
 using an additional signal generator that drives two
variable attenuators acting on both the RF Bragg signals. The
phase noise of the Bragg beams in this configuration was
characterized with a digital phase detector by comparing the beat
note of the two frequency components detected on a photodiode
(placed after the optical fiber) with a reference RF synthesizer
(see appendix).

With the available optical power on the atoms of $P=20$ mW per
beam, the typical optical intensity is $I=250$~mW/cm$^2$ and the
maximum two-photon Rabi frequency estimated for Gaussian pulses at
a detuning $\Delta=8$~GHz is $\Omega = 2\pi\times150$~kHz.
For different Bragg orders the detuning is adjusted to maintain a
high effective Rabi frequency
$\Omega_{\textrm{eff}}=\Omega^n/[(8\omega_r)^{n-1}(n-1)!^2]$ \cite{Giltner1995b}.
The pulse duration is kept larger than $n^{1/6}[\omega_r(n-1)]$
to maintain the losses into other orders negligible \cite{Muller2008}
and thus guarantee high $\pi$-pulse efficiencies.
We set a typical effective Rabi frequency
$\Omega_{\textrm{eff}} = 2\pi\times80$~kHz, with a
$\pi$ pulse duration of $15$~$\mu$s full width at half-maximum (FWHM), corresponding
to a Fourier width larger than the atoms' momentum spread.
At a detuning $\Delta=2.8$ GHz and full
power, we obtain a diffraction efficiency of 50\% for the
4\textsuperscript{th} order.

The experimental sequence is the following:  \textsuperscript{88}Sr atoms
from an atomic beam  produced using a high-efficiency
oven \cite{Schioppo2012} are decelerated in a Zeeman slower and then
trapped and cooled in a two-stage magneto-optical trap (MOT). The
first ``blue'' MOT is realized using the strong
$^{1}S_{0}$--$^{1}P_{1}$ transition  at 461~nm to reach a
temperature of 1~mK. 
The atoms are then further cooled in a ``red'' MOT
operating on the narrow intercombination $^{1}S_{0}$--$^{3}P_{1}$
transition at 689 nm, reaching a final temperature of  $1.2~\mu$K,
with a spatial radial (vertical) size of $300$ $\mu$m (50 $\mu$m) FWHM.
The sequence produces about $2\times10^6$ trapped atoms in $1.5$~s.
A small fraction of the
atoms ($\sim 10^5$) is selected from the MOT and launched upwards with
a sequence of Bragg $\pi$ pulses with a typical duration of 47~$\mu$s FWHM,
up to a total momentum transfer of 40~$\hbar k$.
Even though a single $\pi$ pulse would be
sufficient to isolate the selected atoms from the freely falling cloud after the release from the 
red MOT, a larger number of pulses is applied to increase the
total time of flight up to 150~ms. 
By means of Bragg spectroscopy \cite{Stenger1999} we  estimate a
vertical momentum spread of $1.5~\hbar k$ FWHM for the red MOT,
and $0.2~\hbar k$ for the selected atomic sample, that allows high
fidelity $\pi$ and $\pi/2$ pulses in the interferometer \cite{Szigeti2012}.
Incidentally, in this work we  also performed
preliminary tests of  velocity selection and launch of Sr atoms in a fountain using 
Bloch oscillations in an accelerated vertical optical lattice at 532 nm.
After the launch of the atoms in the fountain, a Mach-Zehnder
interferometer is realized by applying three Bragg pulses in a
$\pi/2$--$\pi$--$\pi/2$ configuration.
As shown in Fig.\ref{fig.apparatus}(b), the first $\pi/2$ pulse coherently splits the atomic wavepacket over two paths separated by $2n\hbar k$. 
Fig.\ref{fig.apparatus}(c) shows an image of the atoms in the two arms of the interferometer
after 30 ms for a 1\textsuperscript{st} order
pulse. The spatial separation
between the two interferometer arms is 600~$\mu$m, that is
about two times larger than the separation induced by near-infrared light in alkali atom interferometers. 
The two  paths in the interferometer are recombined after a time $2T$.
The population in the two output ports is
detected by either absorption imaging or fluorescence collection about
40~ms after the last pulse is applied, when the two
momentum states are sufficiently separated in space.
The interferometer time $T$ is currently limited by the vertical
size of the vacuum chamber (10 cm) which limits the total
time of flight for the atoms in the fountain. 

The number of atoms in the two outputs  $N_{\left|p_{0}\right\rangle }$
and $N_{\left|p_{0}+2n\hbar k\right\rangle }$ is determined by fitting the detected signal with two Gaussian profiles.

They oscillate periodically as a function of the
relative phase $\Phi$ acquired by the atoms in the two arms. The output signal of the interferometer
 $P(\Phi)$ is given by the relative
population:
\begin{equation}\label{eq.contrast}
  P(\Phi)=\frac{N_{\left|p_{0}\right\rangle }}{N_{\left|p_{0}\right\rangle }+N_{\left|p_{0}+2n\hbar k\right\rangle }}=P_{0}+\frac{C}{2}\cos(\Phi)
\end{equation}
where $P_{0}\sim0.5$ is an offset and $C$ is the contrast.
For a vertical Mach-Zehnder Bragg interferometer, the relative phase depends on the
gravity acceleration $g$, the effective laser wave number $2nk$, the interferometer
time $T$ and the optical phase of the Bragg pulses:
\begin{equation}\label{eq.InterfPhase}
  \Phi=n(2kg-\alpha)T^{2}+n(\phi_{1}-2\phi_{2}+\phi_{3})
\end{equation}
where $\alpha = 2\pi\times42.5509$ kHz/ms is a frequency chirping applied to the Bragg  beams
in order to compensate for the varying Doppler shift of the falling atoms,
and $\phi_{i}$ is the relative phase between the two beams for  the $i^{th}$ pulse.

\section{Experimental results and discussion}

\subsection{Contrast}

\begin{figure}[t]\begin{center}
\includegraphics[width=0.47 \textwidth]{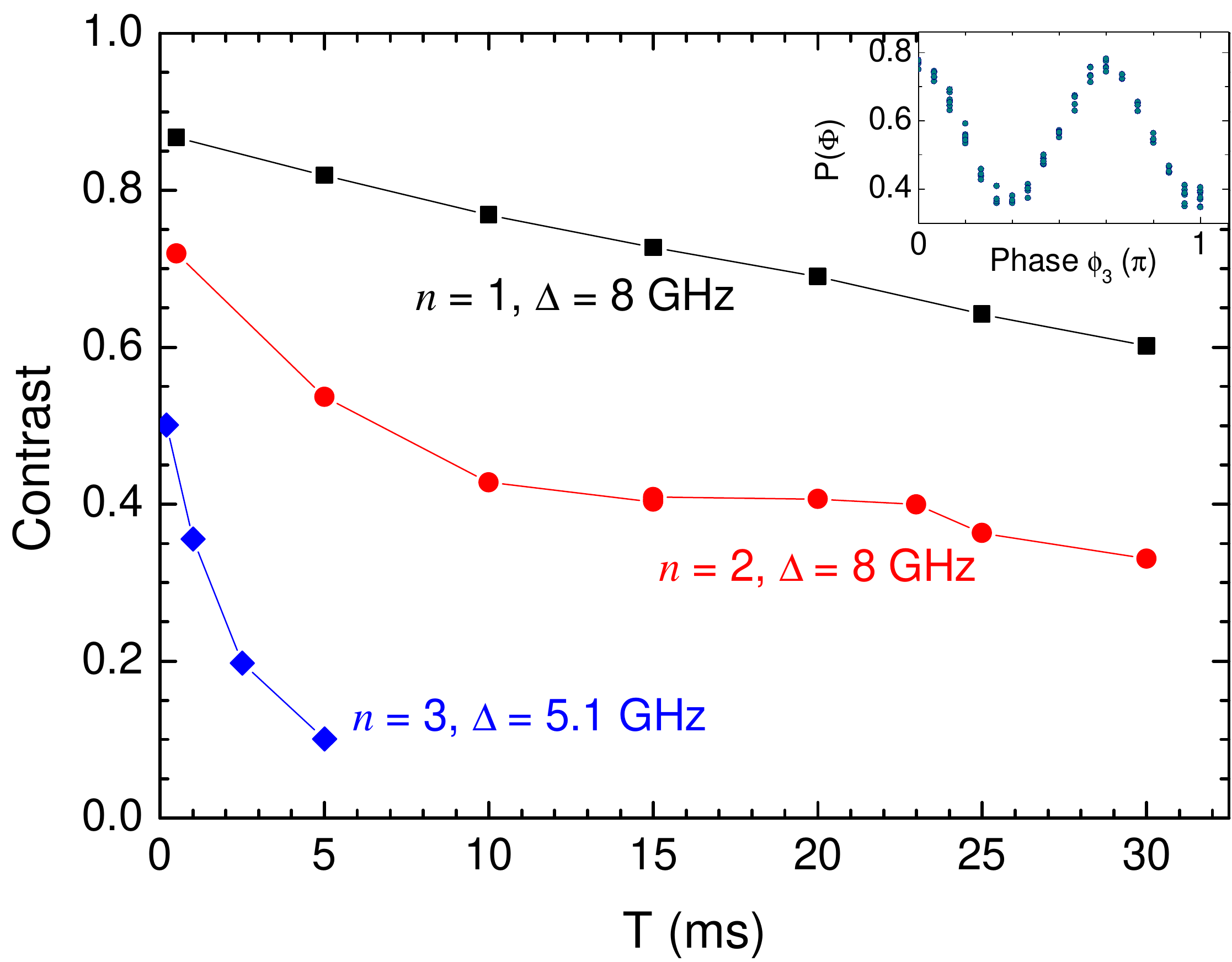}
\caption{Contrast of the interference fringes as a function of 
time $T$ for 1\textsuperscript{st}, 2\textsuperscript{nd} and
3\textsuperscript{rd} order Bragg diffraction with detuning $\Delta$. The inset shows a
typical fringe observed at $T= 0.2$~ms for a 3\textsuperscript{rd}
order Bragg interferometer.} \label{fig.fringecontrast}
\end{center}
\end{figure}

The contrast of the interference fringes, obtained by scanning
the phase $\phi_{3}$ of the last $\pi/2$ pulse, was determined from the values of $P(\Phi)$ between
the 2\textsuperscript{nd} and the 98\textsuperscript{th} percentile
\cite{McDonald2013}.
Fig.~\ref{fig.fringecontrast} shows the values of the observed contrast
for 1\textsuperscript{st}, 2\textsuperscript{nd} and
3\textsuperscript{rd} Bragg order as a function of the interferometer time $T$.
For different orders, the Bragg laser detuning $\Delta$ was chosen
in order to maintain a high Rabi frequency and a low rate of light scattering,
according to the available laser power.
For short interferometer times, the contrast is mainly limited by
the velocity spread along the vertical direction and by the
residual light scattering, which limits the $\pi$ pulse
efficiency.
For long interferometer times, the contrast is mainly limited by the Rabi
frequency inhomogeneity which is due to both the radial expansion of the atomic cloud and the intensity profile imperfections of the Bragg beams
\cite{Muller2008b,McDonald2013b}. 
The sensitivity to this inhomogeneity becomes more critical as the Bragg order n increases because the effective Rabi frequency scales as the $n$\textsuperscript{th} power of the two-photon Rabi frequency.
This  shows that the small sample size and the ultralow temperatures achievable with strontium atoms 
can lead to a high contrast for long interferometer times even with
relatively narrow  Bragg beams. Further improvement in the contrast can be obtained by reducing the probe beam size in order to only interact with the central atoms, for which the Rabi frequency inhomogeneities due to the transverse expansion are smaller. However, in doing this the effect on the sensitivity has to be taken into account. Reducing the interrogation area will reduce the number of interrogated atoms, leading to an increase of the shot noise limit and of detection noise. Therefore, there is a trade-off between contrast gain and noise suppression which has to be optimised in order to really improve the sensitivity of the gravimeter.
Conversely, it is possible to explore geometries where the atoms are guided by a dipole trap along the falling axis. In this scenario the atoms could be forced to remain in the region of maximum intensity of the Bragg beams, ensuring that they all contribute to the interferometer signal.
A technically feasible  improvement by an order of magnitude in the Bragg laser power would allow
us to move further from resonance ($\Delta \sim 600~\Gamma$)
maintaining a sufficiently high Rabi frequency and therefore realize a higher-order  interferometer
as demonstrated for Cs \cite{Muller2008b}.

\begin{figure}[t]\begin{center}
\includegraphics[width=0.47 \textwidth]{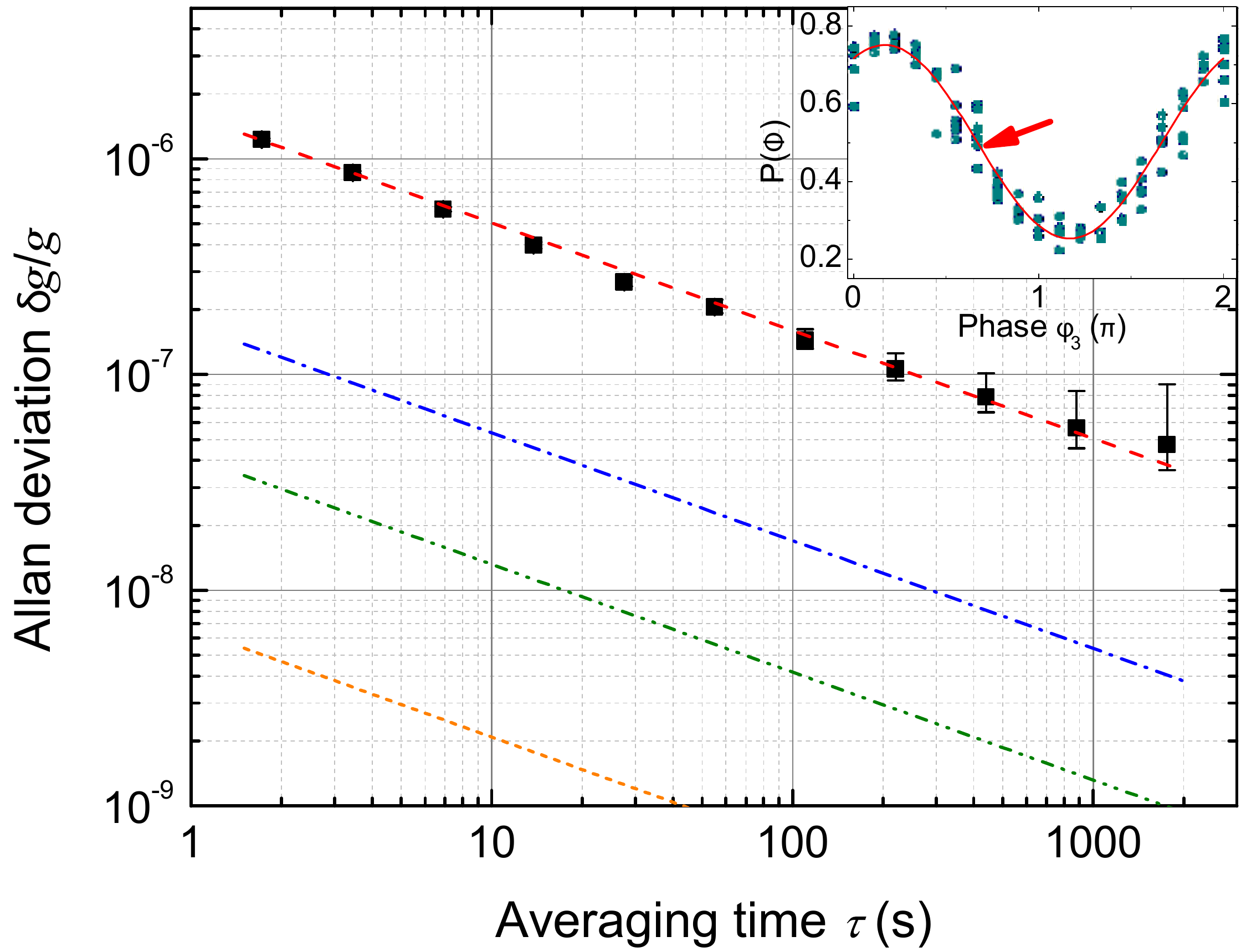}
\caption{Allan deviation of the gravity acceleration measurements
for a 1\textsuperscript{st} order interferometer with a time
$T=30$ ms (black squares). The inset shows the corresponding fringe and the point at which the phase fluctuations are measured. Also shown in the figure are the estimated effects due to 
the residual acceleration noise of the retro-reflection mirror (dash red line),  
the optical phase noise of the Bragg beams (dash dot blue line),  
the intensity noise of the Bragg beams (short-dash orange line) and
the shot noise  ($1\times10^{5}$ atoms,  dash dot dot green line).
} \label{fig.allan}
\end{center}
\end{figure}

\subsection{Sensitivity}

The sensitivity $\delta g/g$ of the interferometer as a gravimeter is determined by measuring
the phase fluctuations $\delta \Phi$ at the slope of the
central fringe:
\begin{equation}\label{eq.sens}
  \frac{\delta g}{g} = \frac{\delta \Phi}{2nkgT^2}.
\end{equation}
The short and long-term sensitivities are
characterized with the Allan deviation. The results for a
1\textsuperscript{st} order interferometer with a time $T=30$~ms
and the estimated effect of the main noise sources are shown in
Fig.~\ref{fig.allan}. The Allan deviation scales as the
inverse-root of the integration time with 
$\delta g/g = 1.5\times10^{-6}$ at 1~s, reaching  $4\times10^{-8}$ at 2000~s. The sensitivity of our interferometer
is presently limited by the residual acceleration of the suspended
retro-reflection mirror.
The estimated phase noise due to the mirror vibrations
is 380 mrad$/\sqrt{\tau}$ where $\tau$ is the averaging time.
The second major noise contribution comes from the optical phase noise
of the Bragg beams which is estimated to be 20 mrad$/\sqrt{\tau}$,
more than one order of magnitude smaller than the vibration noise.
The calculated phase noise arising from intensity fluctuations of the
Bragg laser is 1 mrad$/\sqrt{\tau}$, while other noise sources
such as AC Stark shift effects and Bragg frequency noise are
estimated to give contributions below the $\mu$rad$/\sqrt{\tau}$ level (see
appendix). Finally, the shot noise limit for $10^5$ atoms is $10$ mrad$/\sqrt{\tau}$.

\section{Conclusions and outlook}

In conclusion, we demonstrated LMT Bragg interferometry in a fountain of alkaline-earth atoms for the first time.
The results are mainly limited by technical aspects such as the available laser power, the size of the vacuum cell and residual vibrations; therefore we anticipate a dramatic increase in  performance with the increasing   power of available lasers, a larger chamber to increase the interferometer time and improved isolation from vibrational noise.  A variation on our scheme is the possibility to induce the Bragg transitions using the narrow intercombination line at 689 nm where stable lasers with a higher output power are already available. Moreover,  schemes based on the combination of Bragg diffraction and Bloch oscillations
\cite{Kovachy2010,Muller2009,Charriere2012} might allow superior performances in terms of precision and accuracy thanks to the specific  properties of strontium. Other relevant prospects are the use of ultracold Sr sources \cite{Stellmer2013} and high sensitivity detection schemes beyond the classical limit \cite{Norcia2015}.
%

In order to surpass present limits and take full advantage of the methods and ideas discussed in this paper, we are developing a new apparatus for a large-scale ($\sim$~10~m high) Sr fountain.
Possible fundamental
physics experiments include stringent tests of the Einstein equivalence principle and possible spin-gravity coupling \cite{Poli2011,Tarallo2014}, tests of models of quantum gravity \cite{Amelino2009} and dark matter \cite{Hamilton2015}, new schemes to determine the value of the gravitational constant \cite{Tino2013} and the detection of gravitational waves  \cite{Tino2011}.  Potential applications in geophysics and geodesy can also be envisaged \cite{deAngelis2009}.


In the long term, a space mission based on strontium atoms combining  atom interferometers and transportable  optical clocks \cite{Poli2014} together with a suitable configuration for gravitational wave detection \cite{Yu2011,Graham2013} would enable extremely high precision tests of different fundamental aspects of gravitational physics \cite{Tino2007,Tino2013}.


\section*{Acknowledgments}

We acknowledge support by INFN and the Italian
Ministry of Education, University and Research (MIUR) under the
Progetto Premiale ``Interferometro Atomico'' and by LENS. We also acknowledge
support from the European Union's Seventh Framework Programme
(FP7/2007-2013 grant agreement 250072 - ``iSense'' project and
grant agreement 607493 - ITN ``FACT'' project). We thank G. Rosi
for useful discussions.

\appendix*
\section{Main noise sources in the Sr interferometer}

Here we provide evaluation 
of the main noise sources limiting the sensitivity of the Sr
interferometer. The measured typical power spectral noise
densities (PSD) are also reported. Here, all the equations for the
noise estimation are written for first order Bragg diffraction,
$n=1$. It must be noted that while the estimated phase noise is
proportional to $n$, the sensitivity of the interferometer
($\delta g/g$, as given by the Eq.~\ref{eq.sens}) does not
depend on $n$. For the evaluation, we have considered the
following typical interferometer parameter values: interferometer
time $T=30$ ms, cycle time $T_c$=1.7 s and $\pi/2$ Bragg pulse
duration $\tau_B=10\,\mu$s. In order to maintain simplicity in the
calculations we consider Bragg pulses with a square profile,
although a Gaussian profile is used in the experiment. What is
important for the evaluation of the noise contributions is
actually the pulse area $\tau_B \Omega_R$ and therefore, for
$\tau_B \ll T$, using different pulse shape has a negligible
effect on the result.

\begin{figure}
	\begin{center}
		\includegraphics[width=\linewidth]{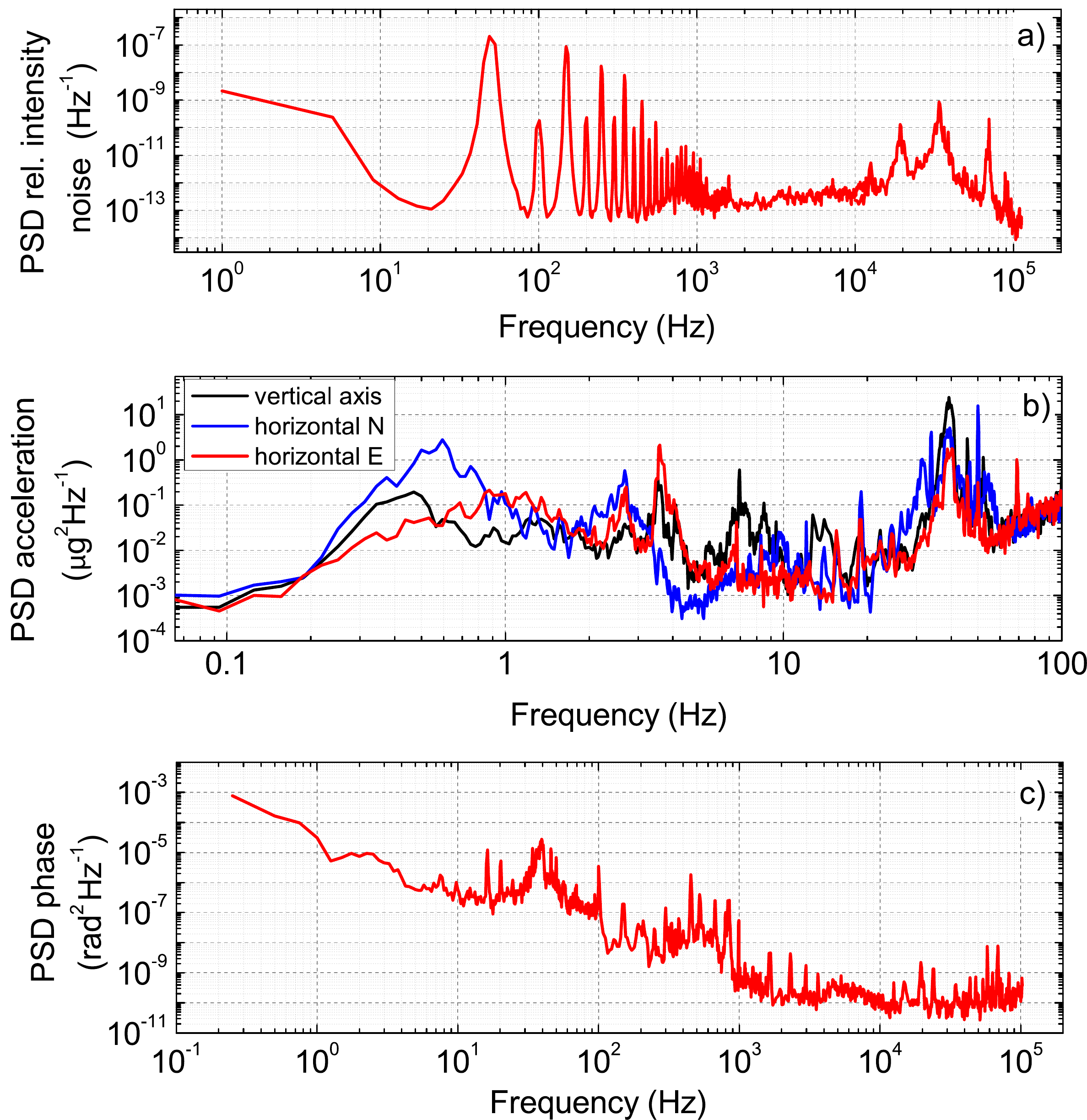}
		\caption{Power spectral densities of main noise sources
		contributing to Sr interferometer phase noise: (a) PSD of relative intensity noise of the 461~nm Bragg laser; (c) PSD of acceleration noise measured on top of the MinusK platform on which is rigidly mounted the retro-reflecting mirror for the Bragg beams, along three orthogonal axes; (b) PSD of phase noise of the Bragg pulses} \label{fig.noise}
	\end{center}
\end{figure}

Fig.\ref{fig.noise}(a) shows the PSD measurement of the Bragg
laser intensity noise. The estimation of phase shift induced on
the interferometer by this noise follows the analysis done for
Raman interferometers (see for example \cite{PetersPhD1998}). The
phase noise, written in terms of its Allan variance, is given by
the following formula:
\begin{equation}
\sigma^2_{\Phi,I}(\tau)=\frac{T_c}{\tau}\int^{+\infty}_{0}S_I(
f)|H_I(f)|^2\mathrm{d} f
\end{equation}
where the sensitivity function $H_I(f)$ is given by
\begin{equation}
|H_I(f)|^2=\frac{\sqrt{3}\pi}{C}\frac{\sin^4(2\pi f T)}{(2\pi f T)^2}
\end{equation}
where $C$ is the contrast of the interferometer. For our selected
values we estimate a $\sigma_{\Phi,I}(\tau)=1$ mrad$/\sqrt{\tau}$.

The PSD measurement of the phase noise $S_\phi(f)$ on the Bragg
beams is presented in Fig.\ref{fig.noise}(b). This has been
characterized through the use of a digital phase and frequency
detector (PFD) by comparing the beatnote of the two Bragg
frequency components, $\omega_1$ and $\omega_2$, to a reference RF
synthesizer. The beatnote is detected on a photodiode placed after
the polarization maintaining fiber (just before the atomic
sample). The integrated phase noise on a time scale of 100 ms is 1
mrad. Under our typical conditions we estimate
$\sigma_{\Phi,\phi}(\tau)=20 $ mrad$/\sqrt{\tau}$, according to
the standard formula \cite{LeGouet2008, Cheinet2008}:
\begin{equation}
\sigma^2_{\Phi,\phi}(\tau)=\frac{T_c^2}{\tau^2}\int^{+\infty}_{0}
\frac{4 \sin^4(\pi f \tau)}{\sin^2(\pi f T_c)} |H_{\phi}(f)|^2
S_\phi(f) \mathrm{d} f
\end{equation}\label{eq.sigma}

where the transfer function $H_{\phi}(f)$ is given, as usual, by
the Fourier transform of the sensitivity function $g(t)$
\begin{equation}
H_{\phi}(f)=2 \pi f \int^{+\infty}_{-\infty}\mathrm{e}^{i 2\pi f
	t}g(t)\mathrm{d} t
\end{equation}
The sensitivity function $g(t)$ for a sequence of three pulses
$\pi/2$---$\pi$---$\pi/2$ of duration
$\tau_B$---2$\tau_B$---$\tau_B$ separated by a time $T$ with Rabi
frequency $\Omega_R$ is :
\begin{displaymath}
g(t) = \left\{ \begin{array}{ll}
0 & \mathrm{for}\,  -T_c/2<t<-T\\
\sin [\Omega_R (t+T)] & \mathrm{for}\,  -T<t<-T+\tau_B\\
1 & \mathrm{for}\,  -T+\tau_B<t<-\tau_B\\
-\sin [\Omega_R t] & \mathrm{for} \, -\tau_B<t<\tau_B\\
-1 & \mathrm{for}\,  \tau_B<t<T-\tau_B\\
\sin [\Omega_R (t-T)] & \mathrm{for}\,  T-\tau_B<t<T\\
0 & \mathrm{for}\,  T<t<T_c/2
\end{array} \right.
\end{displaymath}

Another important contribution to the interferometer noise is
the vibration noise, which is directly coupled to the
upper retro-reflecting mirror for the Bragg beams. The PSD of
acceleration noise $S_a(f)$, measured on top of the supporting
MinusK platform, is shown on Fig.\ref{fig.noise}(c). The
degradation to the Sr interferometer sensitivity due to vibrations
coupled to the retro-reflecting Bragg mirror has been evaluated
with the formula \cite{LeGouet2008, Cheinet2008}:
\begin{equation}
\sigma^2_{\Phi,a}(\tau)=\frac{k^2_{eff}}{\tau}\sum^\infty_{n=1}\frac{|H(2\pi
	n f_c)|^2}{(2\pi n f_c)^4}S_a(2\pi n f_c)
\end{equation}
Here $k_{eff}=2k=4\pi/\lambda$ is the effective wave vector of the
$1^{st}$ order Bragg diffraction with $\lambda=461$ nm. For our
typical vibration noise, we estimated an Allan deviation of
$\sigma_{\Phi,a}(\tau)=380$ mrad$/\sqrt{\tau}$. This contribution
sets the actual limit on our interferometer sensitivity.

Intensity fluctuations of the Bragg laser could, in principle,
induce phase noise through the AC stark shift effect. However, for
Bragg diffraction this effect is reduced in comparison to Raman
interactions, since atoms remain in the same internal state and
only their momentum changes. A residual differential shift comes
from the different detunings for the two momentum states through
the Doppler shift effect. One should therefore still expect a
small contribution to phase noise proportional to the intensity
fluctuation \cite{Altin2013}:
\begin{equation}
\Delta{\phi}_{ac}=\frac{4\delta}{\Delta}\frac{\delta I}{I}
\end{equation}
where ${\delta}$ and ${\Delta}$ are the Bragg resonance
frequency and the Bragg laser detuning respectively, and ${\delta
	I}$ the intensity fluctuation over the interferometer time. With our
typical parameters, we estimate an induced phase noise of
$4\mu$rad per shot, which is negligible compared to other sources of noise.

Finally, the influence of fluctuations of the absolute Bragg laser
wave vector has also been estimated. For this, the frequency
stability of the 461 nm Bragg laser has been characterized through
the beatnote of the Bragg laser against the master cooling laser
at 461 nm. The relative frequency instability at 1 s is 7$\times
10^{-10}$, indicating a relative uncertainty of 7 $\times
10^{-10}$ on $g$, based on the relation:
\begin{equation}
\frac{\Delta g}{\ g}=\frac{\Delta k_{eff}}{\ k_{eff}}=\frac{\Delta \nu}{\ \nu}
\end{equation}

In conclusion, the absolute frequency noise of the Bragg laser is not currently
limiting the performance of the interferometer.


\begin{thebibliography}{48}%
\makeatletter
\providecommand \@ifxundefined [1]{%
 \@ifx{#1\undefined}
}%
\providecommand \@ifnum [1]{%
 \ifnum #1\expandafter \@firstoftwo
 \else \expandafter \@secondoftwo
 \fi
}%
\providecommand \@ifx [1]{%
 \ifx #1\expandafter \@firstoftwo
 \else \expandafter \@secondoftwo
 \fi
}%
\providecommand \natexlab [1]{#1}%
\providecommand \enquote  [1]{``#1''}%
\providecommand \bibnamefont  [1]{#1}%
\providecommand \bibfnamefont [1]{#1}%
\providecommand \citenamefont [1]{#1}%
\providecommand \href@noop [0]{\@secondoftwo}%
\providecommand \href [0]{\begingroup \@sanitize@url \@href}%
\providecommand \@href[1]{\@@startlink{#1}\@@href}%
\providecommand \@@href[1]{\endgroup#1\@@endlink}%
\providecommand \@sanitize@url [0]{\catcode `\\12\catcode `\$12\catcode
  `\&12\catcode `\#12\catcode `\^12\catcode `\_12\catcode `\%12\relax}%
\providecommand \@@startlink[1]{}%
\providecommand \@@endlink[0]{}%
\providecommand \url  [0]{\begingroup\@sanitize@url \@url }%
\providecommand \@url [1]{\endgroup\@href {#1}{\urlprefix }}%
\providecommand \urlprefix  [0]{URL }%
\providecommand \Eprint [0]{\href }%
\providecommand \doibase [0]{http://dx.doi.org/}%
\providecommand \selectlanguage [0]{\@gobble}%
\providecommand \bibinfo  [0]{\@secondoftwo}%
\providecommand \bibfield  [0]{\@secondoftwo}%
\providecommand \translation [1]{[#1]}%
\providecommand \BibitemOpen [0]{}%
\providecommand \bibitemStop [0]{}%
\providecommand \bibitemNoStop [0]{.\EOS\space}%
\providecommand \EOS [0]{\spacefactor3000\relax}%
\providecommand \BibitemShut  [1]{\csname bibitem#1\endcsname}%
\let\auto@bib@innerbib\@empty
\bibitem [{\citenamefont {Tino}\ and\ \citenamefont
  {Kasevich}(2013)}]{Varenna2013}%
  \BibitemOpen
  \bibfield  {author} {\bibinfo {author} {\bibfnamefont {G.~M.}\ \bibnamefont
  {Tino}}\ and\ \bibinfo {author} {\bibfnamefont {M.~A.}\ \bibnamefont
  {Kasevich}},\ }\href@noop {} {\emph {\bibinfo {title} {Atom
  Interferometry}}},\ \bibinfo {edition} {{P}roceedings of the {I}nternational
  {S}chool of {P}hysics ``{E}nrico {F}ermi", {C}ourse {CLXXXVIII}, {V}arenna
  2013}\ ed.\ (\bibinfo  {publisher} {Societ\`a Italiana di Fisica and IOS
  Press},\ \bibinfo {year} {2014})\BibitemShut {NoStop}%
\bibitem [{\citenamefont {Peters}\ \emph {et~al.}(1999)\citenamefont {Peters},
  \citenamefont {Chung},\ and\ \citenamefont {Chu}}]{Peters1999}%
  \BibitemOpen
  \bibfield  {author} {\bibinfo {author} {\bibfnamefont {A.}~\bibnamefont
  {Peters}}, \bibinfo {author} {\bibfnamefont {K.}~\bibnamefont {Chung}}, \
  and\ \bibinfo {author} {\bibfnamefont {S.}~\bibnamefont {Chu}},\ }\href@noop
  {} {\bibfield  {journal} {\bibinfo  {journal} {Nature}\ }\textbf {\bibinfo
  {volume} {400}},\ \bibinfo {pages} {849} (\bibinfo {year}
  {1999})}\BibitemShut {NoStop}%
\bibitem [{\citenamefont {Gillot}\ \emph {et~al.}(2014)\citenamefont {Gillot},
  \citenamefont {Francis}, \citenamefont {Landragin}, \citenamefont
  {Dos~Santos},\ and\ \citenamefont {Merlet}}]{Gillot2014}%
  \BibitemOpen
  \bibfield  {author} {\bibinfo {author} {\bibfnamefont {P.}~\bibnamefont
  {Gillot}}, \bibinfo {author} {\bibfnamefont {O.}~\bibnamefont {Francis}},
  \bibinfo {author} {\bibfnamefont {A.}~\bibnamefont {Landragin}}, \bibinfo
  {author} {\bibfnamefont {F.~P.}\ \bibnamefont {Dos~Santos}}, \ and\ \bibinfo
  {author} {\bibfnamefont {S.}~\bibnamefont {Merlet}},\ }\href@noop {}
  {\bibfield  {journal} {\bibinfo  {journal} {Metrologia}\ }\textbf {\bibinfo
  {volume} {51}},\ \bibinfo {pages} {L15} (\bibinfo {year} {2014})}\BibitemShut
  {NoStop}%
\bibitem [{\citenamefont {McGuirk}\ \emph {et~al.}(2002)\citenamefont
  {McGuirk}, \citenamefont {Foster}, \citenamefont {Fixler}, \citenamefont
  {Snadden},\ and\ \citenamefont {Kasevich}}]{McGuirk2002}%
  \BibitemOpen
  \bibfield  {author} {\bibinfo {author} {\bibfnamefont {J.~M.}\ \bibnamefont
  {McGuirk}}, \bibinfo {author} {\bibfnamefont {G.~T.}\ \bibnamefont {Foster}},
  \bibinfo {author} {\bibfnamefont {J.~B.}\ \bibnamefont {Fixler}}, \bibinfo
  {author} {\bibfnamefont {M.~J.}\ \bibnamefont {Snadden}}, \ and\ \bibinfo
  {author} {\bibfnamefont {M.~A.}\ \bibnamefont {Kasevich}},\ }\href@noop {}
  {\bibfield  {journal} {\bibinfo  {journal} {Phys. Rev. A}\ }\textbf {\bibinfo
  {volume} {65}},\ \bibinfo {pages} {033608} (\bibinfo {year}
  {2002})}\BibitemShut {NoStop}%
\bibitem [{\citenamefont {Sorrentino}\ \emph {et~al.}(2014)\citenamefont
  {Sorrentino}, \citenamefont {Bodart}, \citenamefont {Cacciapuoti},
  \citenamefont {Lien}, \citenamefont {Prevedelli}, \citenamefont {Rosi},
  \citenamefont {Salvi},\ and\ \citenamefont {Tino}}]{Sorrentino2014}%
  \BibitemOpen
  \bibfield  {author} {\bibinfo {author} {\bibfnamefont {F.}~\bibnamefont
  {Sorrentino}}, \bibinfo {author} {\bibfnamefont {Q.}~\bibnamefont {Bodart}},
  \bibinfo {author} {\bibfnamefont {L.}~\bibnamefont {Cacciapuoti}}, \bibinfo
  {author} {\bibfnamefont {Y.-H.}\ \bibnamefont {Lien}}, \bibinfo {author}
  {\bibfnamefont {M.}~\bibnamefont {Prevedelli}}, \bibinfo {author}
  {\bibfnamefont {G.}~\bibnamefont {Rosi}}, \bibinfo {author} {\bibfnamefont
  {L.}~\bibnamefont {Salvi}}, \ and\ \bibinfo {author} {\bibfnamefont {G.~M.}\
  \bibnamefont {Tino}},\ }\href {\doibase 10.1103/PhysRevA.89.023607}
  {\bibfield  {journal} {\bibinfo  {journal} {Phys. Rev. A}\ }\textbf {\bibinfo
  {volume} {89}},\ \bibinfo {pages} {023607} (\bibinfo {year}
  {2014})}\BibitemShut {NoStop}%
\bibitem [{\citenamefont {Rosi}\ \emph {et~al.}(2015)\citenamefont {Rosi},
  \citenamefont {Cacciapuoti}, \citenamefont {Sorrentino}, \citenamefont
  {Menchetti}, \citenamefont {Prevedelli},\ and\ \citenamefont
  {Tino}}]{Rosi2015}%
  \BibitemOpen
  \bibfield  {author} {\bibinfo {author} {\bibfnamefont {G.}~\bibnamefont
  {Rosi}}, \bibinfo {author} {\bibfnamefont {L.}~\bibnamefont {Cacciapuoti}},
  \bibinfo {author} {\bibfnamefont {F.}~\bibnamefont {Sorrentino}}, \bibinfo
  {author} {\bibfnamefont {M.}~\bibnamefont {Menchetti}}, \bibinfo {author}
  {\bibfnamefont {M.}~\bibnamefont {Prevedelli}}, \ and\ \bibinfo {author}
  {\bibfnamefont {G.~M.}\ \bibnamefont {Tino}},\ }\href@noop {} {\bibfield
  {journal} {\bibinfo  {journal} {Phys. Rev. Lett.}\ }\textbf {\bibinfo
  {volume} {114}},\ \bibinfo {pages} {013001} (\bibinfo {year}
  {2015})}\BibitemShut {NoStop}%
\bibitem [{\citenamefont {Rosi}\ \emph {et~al.}(2014)\citenamefont {Rosi},
  \citenamefont {Sorrentino}, \citenamefont {Cacciapuoti}, \citenamefont
  {Prevedelli},\ and\ \citenamefont {Tino}}]{Rosi2014}%
  \BibitemOpen
  \bibfield  {author} {\bibinfo {author} {\bibfnamefont {G.}~\bibnamefont
  {Rosi}}, \bibinfo {author} {\bibfnamefont {F.}~\bibnamefont {Sorrentino}},
  \bibinfo {author} {\bibfnamefont {L.}~\bibnamefont {Cacciapuoti}}, \bibinfo
  {author} {\bibfnamefont {M.}~\bibnamefont {Prevedelli}}, \ and\ \bibinfo
  {author} {\bibfnamefont {G.~M.}\ \bibnamefont {Tino}},\ }\href {\doibase
  {10.1038/nature13433}} {\bibfield  {journal} {\bibinfo  {journal} {{Nature}}\
  }\textbf {\bibinfo {volume} {{510}}},\ \bibinfo {pages} {{518}} (\bibinfo
  {year} {{2014}})}\BibitemShut {NoStop}%
\bibitem [{\citenamefont {Riehle}\ \emph {et~al.}(1991)\citenamefont {Riehle},
  \citenamefont {Kisters}, \citenamefont {Witte}, \citenamefont {Helmcke},\
  and\ \citenamefont {Bord\'e}}]{Riehle1991}%
  \BibitemOpen
  \bibfield  {author} {\bibinfo {author} {\bibfnamefont {F.}~\bibnamefont
  {Riehle}}, \bibinfo {author} {\bibfnamefont {T.}~\bibnamefont {Kisters}},
  \bibinfo {author} {\bibfnamefont {A.}~\bibnamefont {Witte}}, \bibinfo
  {author} {\bibfnamefont {J.}~\bibnamefont {Helmcke}}, \ and\ \bibinfo
  {author} {\bibfnamefont {C.~J.}\ \bibnamefont {Bord\'e}},\ }\href {\doibase
  10.1103/PhysRevLett.67.177} {\bibfield  {journal} {\bibinfo  {journal} {Phys.
  Rev. Lett.}\ }\textbf {\bibinfo {volume} {67}},\ \bibinfo {pages} {177}
  (\bibinfo {year} {1991})}\BibitemShut {NoStop}%
\bibitem [{\citenamefont {Tarallo}\ \emph {et~al.}(2011)\citenamefont
  {Tarallo}, \citenamefont {Poli}, \citenamefont {Schioppo}, \citenamefont
  {Sutyrin},\ and\ \citenamefont {Tino}}]{Tarallo2011}%
  \BibitemOpen
  \bibfield  {author} {\bibinfo {author} {\bibfnamefont {M.~G.}\ \bibnamefont
  {Tarallo}}, \bibinfo {author} {\bibfnamefont {N.}~\bibnamefont {Poli}},
  \bibinfo {author} {\bibfnamefont {M.}~\bibnamefont {Schioppo}}, \bibinfo
  {author} {\bibfnamefont {D.}~\bibnamefont {Sutyrin}}, \ and\ \bibinfo
  {author} {\bibfnamefont {G.~M.}\ \bibnamefont {Tino}},\ }\href@noop {}
  {\bibfield  {journal} {\bibinfo  {journal} {Appl. Phys. B-Lasers Opt.}\
  }\textbf {\bibinfo {volume} {103}},\ \bibinfo {pages} {17} (\bibinfo {year}
  {2011})}\BibitemShut {NoStop}%
\bibitem [{\citenamefont {Graham}\ \emph {et~al.}(2013)\citenamefont {Graham},
  \citenamefont {Hogan}, \citenamefont {Kasevich},\ and\ \citenamefont
  {Rajendran}}]{Graham2013}%
  \BibitemOpen
  \bibfield  {author} {\bibinfo {author} {\bibfnamefont {P.~W.}\ \bibnamefont
  {Graham}}, \bibinfo {author} {\bibfnamefont {J.~M.}\ \bibnamefont {Hogan}},
  \bibinfo {author} {\bibfnamefont {M.~A.}\ \bibnamefont {Kasevich}}, \ and\
  \bibinfo {author} {\bibfnamefont {S.}~\bibnamefont {Rajendran}},\ }\href
  {\doibase 10.1103/PhysRevLett.110.171102} {\bibfield  {journal} {\bibinfo
  {journal} {Phys. Rev. Lett.}\ }\textbf {\bibinfo {volume} {110}},\ \bibinfo
  {pages} {171102} (\bibinfo {year} {2013})}\BibitemShut {NoStop}%
\bibitem [{\citenamefont {Tarallo}\ \emph {et~al.}(2014)\citenamefont
  {Tarallo}, \citenamefont {Mazzoni}, \citenamefont {Poli}, \citenamefont
  {Sutyrin}, \citenamefont {Zhang},\ and\ \citenamefont {Tino}}]{Tarallo2014}%
  \BibitemOpen
  \bibfield  {author} {\bibinfo {author} {\bibfnamefont {M.~G.}\ \bibnamefont
  {Tarallo}}, \bibinfo {author} {\bibfnamefont {T.}~\bibnamefont {Mazzoni}},
  \bibinfo {author} {\bibfnamefont {N.}~\bibnamefont {Poli}}, \bibinfo {author}
  {\bibfnamefont {D.~V.}\ \bibnamefont {Sutyrin}}, \bibinfo {author}
  {\bibfnamefont {X.}~\bibnamefont {Zhang}}, \ and\ \bibinfo {author}
  {\bibfnamefont {G.~M.}\ \bibnamefont {Tino}},\ }\href {\doibase
  10.1103/PhysRevLett.113.023005} {\bibfield  {journal} {\bibinfo  {journal}
  {Phys. Rev. Lett.}\ }\textbf {\bibinfo {volume} {113}},\ \bibinfo {pages}
  {023005} (\bibinfo {year} {2014})}\BibitemShut {NoStop}%
\bibitem [{\citenamefont {Jamison}\ \emph {et~al.}(2014)\citenamefont
  {Jamison}, \citenamefont {Plotkin-Swing},\ and\ \citenamefont
  {Gupta}}]{Jamison2014}%
  \BibitemOpen
  \bibfield  {author} {\bibinfo {author} {\bibfnamefont {A.~O.}\ \bibnamefont
  {Jamison}}, \bibinfo {author} {\bibfnamefont {B.}~\bibnamefont
  {Plotkin-Swing}}, \ and\ \bibinfo {author} {\bibfnamefont {S.}~\bibnamefont
  {Gupta}},\ }\href {\doibase 10.1103/PhysRevA.90.063606} {\bibfield  {journal}
  {\bibinfo  {journal} {Phys. Rev. A}\ }\textbf {\bibinfo {volume} {90}},\
  \bibinfo {pages} {063606} (\bibinfo {year} {2014})}\BibitemShut {NoStop}%
\bibitem [{\citenamefont {Hartwig}\ \emph {et~al.}(2015)\citenamefont
  {Hartwig}, \citenamefont {Abend}, \citenamefont {Schubert}, \citenamefont
  {Schlippert}, \citenamefont {Ahlers}, \citenamefont {Posso-Trujillo},
  \citenamefont {Gaaloul}, \citenamefont {Ertmer},\ and\ \citenamefont
  {Rasel}}]{Hartwig2015}%
  \BibitemOpen
  \bibfield  {author} {\bibinfo {author} {\bibfnamefont {J.}~\bibnamefont
  {Hartwig}}, \bibinfo {author} {\bibfnamefont {S.}~\bibnamefont {Abend}},
  \bibinfo {author} {\bibfnamefont {C.}~\bibnamefont {Schubert}}, \bibinfo
  {author} {\bibfnamefont {D.}~\bibnamefont {Schlippert}}, \bibinfo {author}
  {\bibfnamefont {H.}~\bibnamefont {Ahlers}}, \bibinfo {author} {\bibfnamefont
  {K.}~\bibnamefont {Posso-Trujillo}}, \bibinfo {author} {\bibfnamefont
  {N.}~\bibnamefont {Gaaloul}}, \bibinfo {author} {\bibfnamefont
  {W.}~\bibnamefont {Ertmer}}, \ and\ \bibinfo {author} {\bibfnamefont {E.~M.}\
  \bibnamefont {Rasel}},\ }\href@noop {} {\bibfield  {journal} {\bibinfo
  {journal} {New J. Phys.}\ }\textbf {\bibinfo {volume} {17}},\ \bibinfo
  {pages} {035011} (\bibinfo {year} {2015})}\BibitemShut {NoStop}%
\bibitem [{\citenamefont {Hinkley}\ \emph {et~al.}(2013)\citenamefont
  {Hinkley}, \citenamefont {Sherman}, \citenamefont {Phillips}, \citenamefont
  {Schioppo}, \citenamefont {Lemke}, \citenamefont {Beloy}, \citenamefont
  {Pizzocaro}, \citenamefont {Oates},\ and\ \citenamefont
  {Ludlow}}]{Hinkley2013}%
  \BibitemOpen
  \bibfield  {author} {\bibinfo {author} {\bibfnamefont {N.}~\bibnamefont
  {Hinkley}}, \bibinfo {author} {\bibfnamefont {J.~A.}\ \bibnamefont
  {Sherman}}, \bibinfo {author} {\bibfnamefont {N.~B.}\ \bibnamefont
  {Phillips}}, \bibinfo {author} {\bibfnamefont {M.}~\bibnamefont {Schioppo}},
  \bibinfo {author} {\bibfnamefont {N.~D.}\ \bibnamefont {Lemke}}, \bibinfo
  {author} {\bibfnamefont {K.}~\bibnamefont {Beloy}}, \bibinfo {author}
  {\bibfnamefont {M.}~\bibnamefont {Pizzocaro}}, \bibinfo {author}
  {\bibfnamefont {C.~W.}\ \bibnamefont {Oates}}, \ and\ \bibinfo {author}
  {\bibfnamefont {A.~D.}\ \bibnamefont {Ludlow}},\ }\href@noop {} {\bibfield
  {journal} {\bibinfo  {journal} {Science}\ }\textbf {\bibinfo {volume}
  {341}},\ \bibinfo {pages} {1215} (\bibinfo {year} {2013})}\BibitemShut
  {NoStop}%
\bibitem [{\citenamefont {Ushijima}\ \emph {et~al.}(2015)\citenamefont
  {Ushijima}, \citenamefont {Takamoto}, \citenamefont {Das}, \citenamefont
  {Ohkubo},\ and\ \citenamefont {Katori}}]{Ushijima2015}%
  \BibitemOpen
  \bibfield  {author} {\bibinfo {author} {\bibfnamefont {I.}~\bibnamefont
  {Ushijima}}, \bibinfo {author} {\bibfnamefont {M.}~\bibnamefont {Takamoto}},
  \bibinfo {author} {\bibfnamefont {M.}~\bibnamefont {Das}}, \bibinfo {author}
  {\bibfnamefont {T.}~\bibnamefont {Ohkubo}}, \ and\ \bibinfo {author}
  {\bibfnamefont {H.}~\bibnamefont {Katori}},\ }\href {\doibase
  10.1038/nphoton.2015.5} {\bibfield  {journal} {\bibinfo  {journal} {Nature
  Photon}\ ,\ \bibinfo {pages} {1}} (\bibinfo {year} {2015})}\BibitemShut
  {NoStop}%
\bibitem [{\citenamefont {Bloom}\ \emph {et~al.}(2014)\citenamefont {Bloom},
  \citenamefont {Nicholson}, \citenamefont {Williams}, \citenamefont
  {Campbell}, \citenamefont {Bishof}, \citenamefont {Zhang}, \citenamefont
  {Zhang}, \citenamefont {Bromley},\ and\ \citenamefont {Ye}}]{Bloom2014}%
  \BibitemOpen
  \bibfield  {author} {\bibinfo {author} {\bibfnamefont {B.~J.}\ \bibnamefont
  {Bloom}}, \bibinfo {author} {\bibfnamefont {T.~L.}\ \bibnamefont
  {Nicholson}}, \bibinfo {author} {\bibfnamefont {J.~R.}\ \bibnamefont
  {Williams}}, \bibinfo {author} {\bibfnamefont {S.~L.}\ \bibnamefont
  {Campbell}}, \bibinfo {author} {\bibfnamefont {M.}~\bibnamefont {Bishof}},
  \bibinfo {author} {\bibfnamefont {X.}~\bibnamefont {Zhang}}, \bibinfo
  {author} {\bibfnamefont {W.}~\bibnamefont {Zhang}}, \bibinfo {author}
  {\bibfnamefont {S.~L.}\ \bibnamefont {Bromley}}, \ and\ \bibinfo {author}
  {\bibfnamefont {J.}~\bibnamefont {Ye}},\ }\href@noop {} {\bibfield  {journal}
  {\bibinfo  {journal} {Nature}\ }\textbf {\bibinfo {volume} {506}},\ \bibinfo
  {pages} {71} (\bibinfo {year} {2014})}\BibitemShut {NoStop}%
\bibitem [{\citenamefont {Poli}\ \emph {et~al.}(2013)\citenamefont {Poli},
  \citenamefont {Oates}, \citenamefont {Gill},\ and\ \citenamefont
  {Tino}}]{Poli2013}%
  \BibitemOpen
  \bibfield  {author} {\bibinfo {author} {\bibfnamefont {N.}~\bibnamefont
  {Poli}}, \bibinfo {author} {\bibfnamefont {C.~W.}\ \bibnamefont {Oates}},
  \bibinfo {author} {\bibfnamefont {P.}~\bibnamefont {Gill}}, \ and\ \bibinfo
  {author} {\bibfnamefont {G.~M.}\ \bibnamefont {Tino}},\ }\href@noop {}
  {\bibfield  {journal} {\bibinfo  {journal} {Rivista del Nuovo Cimento}\
  }\textbf {\bibinfo {volume} {36}},\ \bibinfo {pages} {555} (\bibinfo {year}
  {2013})}\BibitemShut {NoStop}%
\bibitem [{\citenamefont {Giltner}\ \emph
  {et~al.}(1995{\natexlab{a}})\citenamefont {Giltner}, \citenamefont
  {McGowan},\ and\ \citenamefont {Lee}}]{Giltner1995}%
  \BibitemOpen
  \bibfield  {author} {\bibinfo {author} {\bibfnamefont {D.~M.}\ \bibnamefont
  {Giltner}}, \bibinfo {author} {\bibfnamefont {R.~W.}\ \bibnamefont
  {McGowan}}, \ and\ \bibinfo {author} {\bibfnamefont {S.~A.}\ \bibnamefont
  {Lee}},\ }\href {\doibase 10.1103/PhysRevLett.75.2638} {\bibfield  {journal}
  {\bibinfo  {journal} {Phys. Rev. Lett.}\ }\textbf {\bibinfo {volume} {75}},\
  \bibinfo {pages} {2638} (\bibinfo {year} {1995}{\natexlab{a}})}\BibitemShut
  {NoStop}%
\bibitem [{\citenamefont {M\"uller}\ \emph
  {et~al.}(2008{\natexlab{a}})\citenamefont {M\"uller}, \citenamefont {Chiow},
  \citenamefont {Long}, \citenamefont {Herrmann},\ and\ \citenamefont
  {Chu}}]{Muller2008b}%
  \BibitemOpen
  \bibfield  {author} {\bibinfo {author} {\bibfnamefont {H.}~\bibnamefont
  {M\"uller}}, \bibinfo {author} {\bibfnamefont {S.-w.}\ \bibnamefont {Chiow}},
  \bibinfo {author} {\bibfnamefont {Q.}~\bibnamefont {Long}}, \bibinfo {author}
  {\bibfnamefont {S.}~\bibnamefont {Herrmann}}, \ and\ \bibinfo {author}
  {\bibfnamefont {S.}~\bibnamefont {Chu}},\ }\href {\doibase
  10.1103/PhysRevLett.100.180405} {\bibfield  {journal} {\bibinfo  {journal}
  {Phys. Rev. Lett.}\ }\textbf {\bibinfo {volume} {100}},\ \bibinfo {pages}
  {180405} (\bibinfo {year} {2008}{\natexlab{a}})}\BibitemShut {NoStop}%
\bibitem [{\citenamefont {Altin}\ \emph {et~al.}(2013)\citenamefont {Altin},
  \citenamefont {Johnsson}, \citenamefont {Negnevitsky}, \citenamefont
  {Dennis}, \citenamefont {Anderson}, \citenamefont {Debs}, \citenamefont
  {Szigeti}, \citenamefont {Hardman}, \citenamefont {Bennetts}, \citenamefont
  {McDonald}, \citenamefont {Turner}, \citenamefont {Close},\ and\
  \citenamefont {Robins}}]{Altin2013}%
  \BibitemOpen
  \bibfield  {author} {\bibinfo {author} {\bibfnamefont {P.~A.}\ \bibnamefont
  {Altin}}, \bibinfo {author} {\bibfnamefont {M.~T.}\ \bibnamefont {Johnsson}},
  \bibinfo {author} {\bibfnamefont {V.}~\bibnamefont {Negnevitsky}}, \bibinfo
  {author} {\bibfnamefont {G.~R.}\ \bibnamefont {Dennis}}, \bibinfo {author}
  {\bibfnamefont {R.~P.}\ \bibnamefont {Anderson}}, \bibinfo {author}
  {\bibfnamefont {J.~E.}\ \bibnamefont {Debs}}, \bibinfo {author}
  {\bibfnamefont {S.~S.}\ \bibnamefont {Szigeti}}, \bibinfo {author}
  {\bibfnamefont {K.~S.}\ \bibnamefont {Hardman}}, \bibinfo {author}
  {\bibfnamefont {S.}~\bibnamefont {Bennetts}}, \bibinfo {author}
  {\bibfnamefont {G.~D.}\ \bibnamefont {McDonald}}, \bibinfo {author}
  {\bibfnamefont {L.~D.}\ \bibnamefont {Turner}}, \bibinfo {author}
  {\bibfnamefont {J.~D.}\ \bibnamefont {Close}}, \ and\ \bibinfo {author}
  {\bibfnamefont {N.~P.}\ \bibnamefont {Robins}},\ }\href@noop {} {\bibfield
  {journal} {\bibinfo  {journal} {New J. Phys.}\ }\textbf {\bibinfo {volume}
  {15}} (\bibinfo {year} {2013})}\BibitemShut {NoStop}%
\bibitem [{\citenamefont {Yu}\ and\ \citenamefont {Tinto}(2011)}]{Yu2011}%
  \BibitemOpen
  \bibfield  {author} {\bibinfo {author} {\bibfnamefont {N.}~\bibnamefont
  {Yu}}\ and\ \bibinfo {author} {\bibfnamefont {M.}~\bibnamefont {Tinto}},\
  }\href {\doibase 10.1007/s10714-010-1055-8} {\bibfield  {journal} {\bibinfo
  {journal} {Gen. Relativ. Gravit.}\ }\textbf {\bibinfo {volume} {43}},\
  \bibinfo {pages} {1943} (\bibinfo {year} {2011})}\BibitemShut {NoStop}%
\bibitem [{\citenamefont {Ferrari}\ \emph {et~al.}(2006)\citenamefont
  {Ferrari}, \citenamefont {Drullinger}, \citenamefont {Poli}, \citenamefont
  {Sorrentino},\ and\ \citenamefont {Tino}}]{Ferrari2006}%
  \BibitemOpen
  \bibfield  {author} {\bibinfo {author} {\bibfnamefont {G.}~\bibnamefont
  {Ferrari}}, \bibinfo {author} {\bibfnamefont {R.~E.}\ \bibnamefont
  {Drullinger}}, \bibinfo {author} {\bibfnamefont {N.}~\bibnamefont {Poli}},
  \bibinfo {author} {\bibfnamefont {F.}~\bibnamefont {Sorrentino}}, \ and\
  \bibinfo {author} {\bibfnamefont {G.~M.}\ \bibnamefont {Tino}},\ }\href@noop
  {} {\bibfield  {journal} {\bibinfo  {journal} {Phys. Rev. A}\ }\textbf
  {\bibinfo {volume} {73}},\ \bibinfo {pages} {023408} (\bibinfo {year}
  {2006})}\BibitemShut {NoStop}%
\bibitem [{\citenamefont {Escobar}\ \emph {et~al.}(2008)\citenamefont
  {Martinez de~Escobar}, \citenamefont {Mickelson}, \citenamefont {Pellegrini},
  \citenamefont {Nagel}, \citenamefont {Traverso}, \citenamefont {Yan},
  \citenamefont {C{\^o}t{\'e}},\ and\ \citenamefont {Killian}}]{Martinez2008}%
  \BibitemOpen
  \bibfield  {author} {\bibinfo {author} {\bibfnamefont {Y.~N.}\
  \bibnamefont {Martinez de~Escobar}}, \bibinfo {author} {\bibfnamefont {P.~G.}~\bibnamefont
  {Mickelson}}, \bibinfo {author} {\bibfnamefont {P.}~\bibnamefont
  {Pellegrini}}, \bibinfo {author} {\bibfnamefont {S.~B.}~\bibnamefont {Nagel}},
  \bibinfo {author} {\bibfnamefont {A.}~\bibnamefont {Traverso}}, \bibinfo
  {author} {\bibfnamefont {M.}~\bibnamefont {Yan}}, \bibinfo {author}
  {\bibfnamefont {R.}~\bibnamefont {C{\^o}t{\'e}}}, \ and\ \bibinfo {author}
  {\bibfnamefont {T.~C.}~\bibnamefont {Killian}},\ }\href {\doibase
  10.1103/PhysRevA.78.062708} {\bibfield  {journal} {\bibinfo  {journal}
  {Phys. Rev. A}\ }\textbf {\bibinfo {volume} {78}},\ \bibinfo {pages}
  {062708} (\bibinfo {year} {2008})}\BibitemShut {NoStop}%
\bibitem [{\citenamefont {Stein}\ \emph {et~al.}(2010)\citenamefont {Stein},
  \citenamefont {Kn{\"o}ckel},\ and\ \citenamefont {Tiemann}}]{Stein2010}%
  \BibitemOpen
  \bibfield  {author} {\bibinfo {author} {\bibfnamefont {A.}~\bibnamefont
  {Stein}}, \bibinfo {author} {\bibfnamefont {H.}~\bibnamefont {Kn{\"o}ckel}},
  \ and\ \bibinfo {author} {\bibfnamefont {E.}~\bibnamefont {Tiemann}},\ }\href
  {\doibase 10.1140/epjd/e2010-00058-y} {\bibfield  {journal} {\bibinfo
  {journal} {Eur. Phys. J. D}\ }\textbf {\bibinfo {volume} {57}},\ \bibinfo
  {pages} {171} (\bibinfo {year} {2010})}\BibitemShut {NoStop}%
\bibitem [{\citenamefont {Poli}\ \emph {et~al.}(2011)\citenamefont {Poli},
  \citenamefont {Wang}, \citenamefont {Tarallo}, \citenamefont {Alberti},
  \citenamefont {Prevedelli},\ and\ \citenamefont {Tino}}]{Poli2011}%
  \BibitemOpen
  \bibfield  {author} {\bibinfo {author} {\bibfnamefont {N.}~\bibnamefont
  {Poli}}, \bibinfo {author} {\bibfnamefont {F.-Y.}\ \bibnamefont {Wang}},
  \bibinfo {author} {\bibfnamefont {M.~G.}\ \bibnamefont {Tarallo}}, \bibinfo
  {author} {\bibfnamefont {A.}~\bibnamefont {Alberti}}, \bibinfo {author}
  {\bibfnamefont {M.}~\bibnamefont {Prevedelli}}, \ and\ \bibinfo {author}
  {\bibfnamefont {G.~M.}\ \bibnamefont {Tino}},\ }\href {\doibase
  10.1103/PhysRevLett.106.038501} {\bibfield  {journal} {\bibinfo  {journal}
  {Phys. Rev. Lett.}\ }\textbf {\bibinfo {volume} {106}},\ \bibinfo {pages}
  {038501} (\bibinfo {year} {2011})}\BibitemShut {NoStop}%
\bibitem [{\citenamefont {M\"uller}\ \emph
  {et~al.}(2008{\natexlab{b}})\citenamefont {M\"uller}, \citenamefont {Chiow},\
  and\ \citenamefont {Chu}}]{Muller2008}%
  \BibitemOpen
  \bibfield  {author} {\bibinfo {author} {\bibfnamefont {H.}~\bibnamefont
  {M\"uller}}, \bibinfo {author} {\bibfnamefont {S.-w.}\ \bibnamefont {Chiow}},
  \ and\ \bibinfo {author} {\bibfnamefont {S.}~\bibnamefont {Chu}},\ }\href
  {\doibase 10.1103/PhysRevA.77.023609} {\bibfield  {journal} {\bibinfo
  {journal} {Phys. Rev. A}\ }\textbf {\bibinfo {volume} {77}},\ \bibinfo
  {pages} {023609} (\bibinfo {year} {2008}{\natexlab{b}})}\BibitemShut
  {NoStop}%
\bibitem [{\citenamefont {Giltner}\ \emph
  {et~al.}(1995{\natexlab{b}})\citenamefont {Giltner}, \citenamefont
  {McGowan},\ and\ \citenamefont {Lee}}]{Giltner1995b}%
  \BibitemOpen
  \bibfield  {author} {\bibinfo {author} {\bibfnamefont {D.~M.}\ \bibnamefont
  {Giltner}}, \bibinfo {author} {\bibfnamefont {R.~W.}\ \bibnamefont
  {McGowan}}, \ and\ \bibinfo {author} {\bibfnamefont {S.~A.}\ \bibnamefont
  {Lee}},\ }\href {\doibase 10.1103/PhysRevA.52.3966} {\bibfield  {journal}
  {\bibinfo  {journal} {Phys. Rev. A}\ }\textbf {\bibinfo {volume} {52}},\
  \bibinfo {pages} {3966} (\bibinfo {year} {1995}{\natexlab{b}})}\BibitemShut
  {NoStop}%
\bibitem [{\citenamefont {Schioppo}\ \emph {et~al.}(2012)\citenamefont
  {Schioppo}, \citenamefont {Poli}, \citenamefont {Prevedelli}, \citenamefont
  {Falke}, \citenamefont {Lisdat}, \citenamefont {Sterr},\ and\ \citenamefont
  {Tino}}]{Schioppo2012}%
  \BibitemOpen
  \bibfield  {author} {\bibinfo {author} {\bibfnamefont {M.}~\bibnamefont
  {Schioppo}}, \bibinfo {author} {\bibfnamefont {N.}~\bibnamefont {Poli}},
  \bibinfo {author} {\bibfnamefont {M.}~\bibnamefont {Prevedelli}}, \bibinfo
  {author} {\bibfnamefont {S.}~\bibnamefont {Falke}}, \bibinfo {author}
  {\bibfnamefont {C.}~\bibnamefont {Lisdat}}, \bibinfo {author} {\bibfnamefont
  {U.}~\bibnamefont {Sterr}}, \ and\ \bibinfo {author} {\bibfnamefont {G.~M.}\
  \bibnamefont {Tino}},\ }\href@noop {} {\bibfield  {journal} {\bibinfo
  {journal} {Rev. Sci. Instrum.}\ }\textbf {\bibinfo {volume} {83}} (\bibinfo
  {year} {2012})}\BibitemShut {NoStop}%
\bibitem [{\citenamefont {Stenger}\ \emph {et~al.}(1999)\citenamefont
  {Stenger}, \citenamefont {Inouye}, \citenamefont {Chikkatur}, \citenamefont
  {Stamper-Kurn}, \citenamefont {Pritchard},\ and\ \citenamefont
  {Ketterle}}]{Stenger1999}%
  \BibitemOpen
  \bibfield  {author} {\bibinfo {author} {\bibfnamefont {J.}~\bibnamefont
  {Stenger}}, \bibinfo {author} {\bibfnamefont {S.}~\bibnamefont {Inouye}},
  \bibinfo {author} {\bibfnamefont {A.~P.}\ \bibnamefont {Chikkatur}}, \bibinfo
  {author} {\bibfnamefont {D.~M.}\ \bibnamefont {Stamper-Kurn}}, \bibinfo
  {author} {\bibfnamefont {D.~E.}\ \bibnamefont {Pritchard}}, \ and\ \bibinfo
  {author} {\bibfnamefont {W.}~\bibnamefont {Ketterle}},\ }\href@noop {}
  {\bibfield  {journal} {\bibinfo  {journal} {Phys. Rev. Lett.}\ }\textbf
  {\bibinfo {volume} {82}},\ \bibinfo {pages} {4569} (\bibinfo {year}
  {1999})}\BibitemShut {NoStop}%
\bibitem [{\citenamefont {Szigeti}\ \emph {et~al.}(2012)\citenamefont
  {Szigeti}, \citenamefont {Debs}, \citenamefont {Hope}, \citenamefont
  {Robins},\ and\ \citenamefont {Close}}]{Szigeti2012}%
  \BibitemOpen
  \bibfield  {author} {\bibinfo {author} {\bibfnamefont {S.~S.}\ \bibnamefont
  {Szigeti}}, \bibinfo {author} {\bibfnamefont {J.~E.}\ \bibnamefont {Debs}},
  \bibinfo {author} {\bibfnamefont {J.~J.}\ \bibnamefont {Hope}}, \bibinfo
  {author} {\bibfnamefont {N.~P.}\ \bibnamefont {Robins}}, \ and\ \bibinfo
  {author} {\bibfnamefont {J.~D.}\ \bibnamefont {Close}},\ }\href@noop {}
  {\bibfield  {journal} {\bibinfo  {journal} {New J. Phys.}\ }\textbf {\bibinfo
  {volume} {14}},\ \bibinfo {pages} {023009} (\bibinfo {year}
  {2012})}\BibitemShut {NoStop}%
\bibitem [{\citenamefont {McDonald}\ \emph
  {et~al.}(2013{\natexlab{a}})\citenamefont {McDonald}, \citenamefont {Keal},
  \citenamefont {Altin}, \citenamefont {Debs}, \citenamefont {Bennetts},
  \citenamefont {Kuhn}, \citenamefont {Hardman}, \citenamefont {Johnsson},
  \citenamefont {Close},\ and\ \citenamefont {Robins}}]{McDonald2013}%
  \BibitemOpen
  \bibfield  {author} {\bibinfo {author} {\bibfnamefont {G.~D.}\ \bibnamefont
  {McDonald}}, \bibinfo {author} {\bibfnamefont {H.}~\bibnamefont {Keal}},
  \bibinfo {author} {\bibfnamefont {P.~A.}\ \bibnamefont {Altin}}, \bibinfo
  {author} {\bibfnamefont {J.~E.}\ \bibnamefont {Debs}}, \bibinfo {author}
  {\bibfnamefont {S.}~\bibnamefont {Bennetts}}, \bibinfo {author}
  {\bibfnamefont {C.~C.~N.}\ \bibnamefont {Kuhn}}, \bibinfo {author}
  {\bibfnamefont {K.~S.}\ \bibnamefont {Hardman}}, \bibinfo {author}
  {\bibfnamefont {M.~T.}\ \bibnamefont {Johnsson}}, \bibinfo {author}
  {\bibfnamefont {J.~D.}\ \bibnamefont {Close}}, \ and\ \bibinfo {author}
  {\bibfnamefont {N.~P.}\ \bibnamefont {Robins}},\ }\href {\doibase
  10.1103/PhysRevA.87.013632} {\bibfield  {journal} {\bibinfo  {journal} {Phys.
  Rev. A}\ }\textbf {\bibinfo {volume} {87}},\ \bibinfo {pages} {013632}
  (\bibinfo {year} {2013}{\natexlab{a}})}\BibitemShut {NoStop}%
\bibitem [{\citenamefont {McDonald}\ \emph
  {et~al.}(2013{\natexlab{b}})\citenamefont {McDonald}, \citenamefont {Kuhn},
  \citenamefont {Bennetts}, \citenamefont {Debs}, \citenamefont {Hardman},
  \citenamefont {Johnsson}, \citenamefont {Close},\ and\ \citenamefont
  {Robins}}]{McDonald2013b}%
  \BibitemOpen
  \bibfield  {author} {\bibinfo {author} {\bibfnamefont {G.~D.}\ \bibnamefont
  {McDonald}}, \bibinfo {author} {\bibfnamefont {C.~C.~N.}\ \bibnamefont
  {Kuhn}}, \bibinfo {author} {\bibfnamefont {S.}~\bibnamefont {Bennetts}},
  \bibinfo {author} {\bibfnamefont {J.~E.}\ \bibnamefont {Debs}}, \bibinfo
  {author} {\bibfnamefont {K.~S.}\ \bibnamefont {Hardman}}, \bibinfo {author}
  {\bibfnamefont {M.}~\bibnamefont {Johnsson}}, \bibinfo {author}
  {\bibfnamefont {J.~D.}\ \bibnamefont {Close}}, \ and\ \bibinfo {author}
  {\bibfnamefont {N.~P.}\ \bibnamefont {Robins}},\ }\href {\doibase
  10.1103/PhysRevA.88.053620} {\bibfield  {journal} {\bibinfo  {journal} {Phys.
  Rev. A}\ }\textbf {\bibinfo {volume} {88}},\ \bibinfo {pages} {053620}
  (\bibinfo {year} {2013}{\natexlab{b}})}\BibitemShut {NoStop}%
\bibitem [{\citenamefont {Kovachy}\ \emph {et~al.}(2010)\citenamefont
  {Kovachy}, \citenamefont {Hogan}, \citenamefont {Johnson},\ and\
  \citenamefont {Kasevich}}]{Kovachy2010}%
  \BibitemOpen
  \bibfield  {author} {\bibinfo {author} {\bibfnamefont {T.}~\bibnamefont
  {Kovachy}}, \bibinfo {author} {\bibfnamefont {J.~M.}\ \bibnamefont {Hogan}},
  \bibinfo {author} {\bibfnamefont {D.~M.~S.}\ \bibnamefont {Johnson}}, \ and\
  \bibinfo {author} {\bibfnamefont {M.~A.}\ \bibnamefont {Kasevich}},\ }\href
  {\doibase 10.1103/PhysRevA.82.013638} {\bibfield  {journal} {\bibinfo
  {journal} {Phys. Rev. A}\ }\textbf {\bibinfo {volume} {82}},\ \bibinfo
  {pages} {013638} (\bibinfo {year} {2010})}\BibitemShut {NoStop}%
\bibitem [{\citenamefont {M\"uller}\ \emph {et~al.}(2009)\citenamefont
  {M\"uller}, \citenamefont {Chiow}, \citenamefont {Herrmann},\ and\
  \citenamefont {Chu}}]{Muller2009}%
  \BibitemOpen
  \bibfield  {author} {\bibinfo {author} {\bibfnamefont {H.}~\bibnamefont
  {M\"uller}}, \bibinfo {author} {\bibfnamefont {S.-w.}\ \bibnamefont {Chiow}},
  \bibinfo {author} {\bibfnamefont {S.}~\bibnamefont {Herrmann}}, \ and\
  \bibinfo {author} {\bibfnamefont {S.}~\bibnamefont {Chu}},\ }\href {\doibase
  10.1103/PhysRevLett.102.240403} {\bibfield  {journal} {\bibinfo  {journal}
  {Phys. Rev. Lett.}\ }\textbf {\bibinfo {volume} {102}},\ \bibinfo {pages}
  {240403} (\bibinfo {year} {2009})}\BibitemShut {NoStop}%
\bibitem [{\citenamefont {Charri\`ere}\ \emph {et~al.}(2012)\citenamefont
  {Charri\`ere}, \citenamefont {Cadoret}, \citenamefont {Zahzam}, \citenamefont
  {Bidel},\ and\ \citenamefont {Bresson}}]{Charriere2012}%
  \BibitemOpen
  \bibfield  {author} {\bibinfo {author} {\bibfnamefont {R.}~\bibnamefont
  {Charri\`ere}}, \bibinfo {author} {\bibfnamefont {M.}~\bibnamefont
  {Cadoret}}, \bibinfo {author} {\bibfnamefont {N.}~\bibnamefont {Zahzam}},
  \bibinfo {author} {\bibfnamefont {Y.}~\bibnamefont {Bidel}}, \ and\ \bibinfo
  {author} {\bibfnamefont {A.}~\bibnamefont {Bresson}},\ }\href {\doibase
  10.1103/PhysRevA.85.013639} {\bibfield  {journal} {\bibinfo  {journal} {Phys.
  Rev. A}\ }\textbf {\bibinfo {volume} {85}},\ \bibinfo {pages} {013639}
  (\bibinfo {year} {2012})}\BibitemShut {NoStop}%
\bibitem [{\citenamefont {Stellmer}\ \emph {et~al.}(2013)\citenamefont
  {Stellmer}, \citenamefont {Grimm},\ and\ \citenamefont
  {Schreck}}]{Stellmer2013}%
  \BibitemOpen
  \bibfield  {author} {\bibinfo {author} {\bibfnamefont {S.}~\bibnamefont
  {Stellmer}}, \bibinfo {author} {\bibfnamefont {R.}~\bibnamefont {Grimm}}, \
  and\ \bibinfo {author} {\bibfnamefont {F.}~\bibnamefont {Schreck}},\ }\href
  {\doibase 10.1103/PhysRevA.87.013611} {\bibfield  {journal} {\bibinfo
  {journal} {Phys. Rev. A}\ }\textbf {\bibinfo {volume} {87}},\ \bibinfo
  {pages} {013611} (\bibinfo {year} {2013})}\BibitemShut {NoStop}%
\bibitem [{\citenamefont {Norcia}\ and\ \citenamefont
  {Thompson}(2015)}]{Norcia2015}%
  \BibitemOpen
  \bibfield  {author} {\bibinfo {author} {\bibfnamefont {M.~A.}\ \bibnamefont
  {Norcia}}\ and\ \bibinfo {author} {\bibfnamefont {J.~K.}\ \bibnamefont
  {Thompson}},\ }\href {http://arxiv.org/abs/1506.02297v1} {\bibfield
  {journal} {\bibinfo  {journal} {arXiv}\ }\textbf {\bibinfo {volume}
  {physics.atom-ph}} (\bibinfo {year} {2015})},\ \Eprint
  {http://arxiv.org/abs/1506.02297v1} {1506.02297v1} \BibitemShut {NoStop}%
\bibitem [{\citenamefont {Amelino-Camelia}\ \emph {et~al.}(2009)\citenamefont
  {Amelino-Camelia}, \citenamefont {L\"ammerzahl}, \citenamefont {Mercati},\
  and\ \citenamefont {Tino}}]{Amelino2009}%
  \BibitemOpen
  \bibfield  {author} {\bibinfo {author} {\bibfnamefont {G.}~\bibnamefont
  {Amelino-Camelia}}, \bibinfo {author} {\bibfnamefont {C.}~\bibnamefont
  {L\"ammerzahl}}, \bibinfo {author} {\bibfnamefont {F.}~\bibnamefont
  {Mercati}}, \ and\ \bibinfo {author} {\bibfnamefont {G.~M.}\ \bibnamefont
  {Tino}},\ }\href@noop {} {\bibfield  {journal} {\bibinfo  {journal} {Phys.
  Rev. Lett.}\ }\textbf {\bibinfo {volume} {103}},\ \bibinfo {pages} {171302}
  (\bibinfo {year} {2009})}\BibitemShut {NoStop}%
\bibitem [{\citenamefont {Hamilton}\ \emph {et~al.}(2015)\citenamefont
  {Hamilton}, \citenamefont {Jaffe}, \citenamefont {Haslinger}, \citenamefont
  {Simmons}, \citenamefont {M\"uller},\ and\ \citenamefont
  {Khoury}}]{Hamilton2015}%
  \BibitemOpen
  \bibfield  {author} {\bibinfo {author} {\bibfnamefont {P.}~\bibnamefont
  {Hamilton}}, \bibinfo {author} {\bibfnamefont {M.}~\bibnamefont {Jaffe}},
  \bibinfo {author} {\bibfnamefont {P.}~\bibnamefont {Haslinger}}, \bibinfo
  {author} {\bibfnamefont {Q.}~\bibnamefont {Simmons}}, \bibinfo {author}
  {\bibfnamefont {H.}~\bibnamefont {M\"uller}}, \ and\ \bibinfo {author}
  {\bibfnamefont {J.}~\bibnamefont {Khoury}},\ }\href@noop {} {\bibfield
  {journal} {\bibinfo  {journal} {arXiv:1502.03888}\ } (\bibinfo {year}
  {2015})}\BibitemShut {NoStop}%
\bibitem [{\citenamefont {Tino}(2013)}]{Tino2013}%
  \BibitemOpen
  \bibfield  {author} {\bibinfo {author} {\bibfnamefont {G.~M.}\ \bibnamefont
  {Tino}},\ }\href@noop {} {\bibfield  {journal} {\bibinfo  {journal} {in [1]},\ p.\ \bibinfo {pages} {457}}
	}\BibitemShut {NoStop}%
\bibitem [{Tin(2011)}]{Tino2011}%
  \BibitemOpen
  Special\ issue\ in\ \href@noop {} {\emph {\bibinfo {booktitle} {Gen. Relativ. Gravit.}}},\
  Vol.~\bibinfo {volume} {43},\ \bibinfo {editor} {edited by\ \bibinfo {editor}
  {\bibfnamefont {G.~M.}\ \bibnamefont {Tino}}, \bibinfo {editor}
  {\bibfnamefont {F.}~\bibnamefont {Vetrano}}, \ and\ \bibinfo {editor}
  {\bibfnamefont {C.}~\bibnamefont {L{\"a}mmerzahl}}},\ p.\ \bibinfo {pages} {1901}\ (\bibinfo {year}{2011})\BibitemShut {NoStop}%
\bibitem [{\citenamefont {de~Angelis}\ \emph {et~al.}(2009)\citenamefont
  {de~Angelis}, \citenamefont {Bertoldi}, \citenamefont {Cacciapuoti},
  \citenamefont {Giorgini}, \citenamefont {Lamporesi}, \citenamefont
  {Prevedelli}, \citenamefont {Saccorotti}, \citenamefont {Sorrentino},\ and\
  \citenamefont {Tino}}]{deAngelis2009}%
  \BibitemOpen
  \bibfield  {author} {\bibinfo {author} {\bibfnamefont {M.}~\bibnamefont
  {de~Angelis}}, \bibinfo {author} {\bibfnamefont {A.}~\bibnamefont
  {Bertoldi}}, \bibinfo {author} {\bibfnamefont {L.}~\bibnamefont
  {Cacciapuoti}}, \bibinfo {author} {\bibfnamefont {A.}~\bibnamefont
  {Giorgini}}, \bibinfo {author} {\bibfnamefont {G.}~\bibnamefont {Lamporesi}},
  \bibinfo {author} {\bibfnamefont {M.}~\bibnamefont {Prevedelli}}, \bibinfo
  {author} {\bibfnamefont {G.}~\bibnamefont {Saccorotti}}, \bibinfo {author}
  {\bibfnamefont {F.}~\bibnamefont {Sorrentino}}, \ and\ \bibinfo {author}
  {\bibfnamefont {G.}~\bibnamefont {Tino}},\ }\href@noop {} {\bibfield
  {journal} {\bibinfo  {journal} {Meas. Sci. Technol}\ }\textbf {\bibinfo
  {volume} {20}},\ \bibinfo {pages} {022001} (\bibinfo {year}
  {2009})}\BibitemShut {NoStop}%
\bibitem [{\citenamefont {Poli}\ \emph {et~al.}(2014)\citenamefont {Poli},
  \citenamefont {Schioppo}, \citenamefont {Vogt}, \citenamefont {Falke},
  \citenamefont {Sterr}, \citenamefont {Lisdat},\ and\ \citenamefont
  {Tino}}]{Poli2014}%
  \BibitemOpen
  \bibfield  {author} {\bibinfo {author} {\bibfnamefont {N.}~\bibnamefont
  {Poli}}, \bibinfo {author} {\bibfnamefont {M.}~\bibnamefont {Schioppo}},
  \bibinfo {author} {\bibfnamefont {S.}~\bibnamefont {Vogt}}, \bibinfo {author}
  {\bibfnamefont {S.}~\bibnamefont {Falke}}, \bibinfo {author} {\bibfnamefont
  {U.}~\bibnamefont {Sterr}}, \bibinfo {author} {\bibfnamefont
  {Ch.}~\bibnamefont {Lisdat}}, \ and\ \bibinfo {author} {\bibfnamefont {G.~M.}\
  \bibnamefont {Tino}},\ }\href@noop {} {\bibfield  {journal} {\bibinfo
  {journal} {Applied Physics B}\ }\textbf {\bibinfo {volume} {117(4)}},\
  \bibinfo {pages} {1107} (\bibinfo {year} {2014})}\BibitemShut {NoStop}%
\bibitem [{\citenamefont {Tino}\ \emph {et~al.}(2007)\citenamefont {Tino},
  \citenamefont {Cacciapuoti}, \citenamefont {Bongs}, \citenamefont
  {Bord{\'e}}, \citenamefont {Bouyer}, \citenamefont {Dittus}, \citenamefont
  {Ertmer}, \citenamefont {G{\"o}rlitz}, \citenamefont {Inguscio},
  \citenamefont {Landragin}, \citenamefont {Lemonde}, \citenamefont
  {Lammerzahl}, \citenamefont {Peters}, \citenamefont {Rasel}, \citenamefont
  {Reichel}, \citenamefont {Salomon}, \citenamefont {Schiller}, \citenamefont
  {Schleich}, \citenamefont {Sengstock}, \citenamefont {Sterr},\ and\
  \citenamefont {Wilkens}}]{Tino2007}%
  \BibitemOpen
  \bibfield  {author} {\bibinfo {author} {\bibfnamefont {G.~M.}\ \bibnamefont
  {Tino}}, \bibinfo {author} {\bibfnamefont {L.}~\bibnamefont {Cacciapuoti}},
  \bibinfo {author} {\bibfnamefont {K.}~\bibnamefont {Bongs}}, \bibinfo
  {author} {\bibfnamefont {C.~J.}\ \bibnamefont {Bord{\'e}}}, \bibinfo {author}
  {\bibfnamefont {P.}~\bibnamefont {Bouyer}}, \bibinfo {author} {\bibfnamefont
  {H.}~\bibnamefont {Dittus}}, \bibinfo {author} {\bibfnamefont
  {W.}~\bibnamefont {Ertmer}}, \bibinfo {author} {\bibfnamefont
  {A.}~\bibnamefont {G{\"o}rlitz}}, \bibinfo {author} {\bibfnamefont
  {M.}~\bibnamefont {Inguscio}}, \bibinfo {author} {\bibfnamefont
  {A.}~\bibnamefont {Landragin}}, \bibinfo {author} {\bibfnamefont
  {P.}~\bibnamefont {Lemonde}}, \bibinfo {author} {\bibfnamefont
  {C.}~\bibnamefont {Lammerzahl}}, \bibinfo {author} {\bibfnamefont
  {A.}~\bibnamefont {Peters}}, \bibinfo {author} {\bibfnamefont
  {E.}~\bibnamefont {Rasel}}, \bibinfo {author} {\bibfnamefont
  {J.}~\bibnamefont {Reichel}}, \bibinfo {author} {\bibfnamefont
  {C.}~\bibnamefont {Salomon}}, \bibinfo {author} {\bibfnamefont
  {S.}~\bibnamefont {Schiller}}, \bibinfo {author} {\bibfnamefont
  {W.}~\bibnamefont {Schleich}}, \bibinfo {author} {\bibfnamefont
  {K.}~\bibnamefont {Sengstock}}, \bibinfo {author} {\bibfnamefont
  {U.}~\bibnamefont {Sterr}}, \ and\ \bibinfo {author} {\bibfnamefont
  {M.}~\bibnamefont {Wilkens}},\ }\href {\doibase
  10.1016/j.nuclphysbps.2006.12.061} {\bibfield  {journal} {\bibinfo  {journal}
  {Nuclear Physics B (Proceedings Supplements)}\ }\textbf {\bibinfo {volume}
  {166}},\ \bibinfo{pages}{159}(\bibinfo{year}{2007})}\BibitemShut{NoStop}%
\bibitem{PetersPhD1998} A. Peters, Ph.D thesis, \emph{High Precision
	Gravity measurements using atom interferometry}, Stanford
University (1998).
\bibitem{LeGouet2008}  J. Le Gou{\"e}t, T.E. Mehlst{\"a}ubler, J. Kim, S. Merlet,
A. Clairon, A. Landragin, F. Pereira Dos Santos, Appl.
Phys. B, {\bf 92}, 133 (2008).
\bibitem{Cheinet2008} P. Cheinet, B. Canuel, F. Pereira Dos Santos, A. Gauguet,
F. Leduc, A. Landragin, 
IEEE Trans. Instrum. Meas. {\bf 57}, 1141 (2008).
\bibitem{Altin2013} P. A. Altin, M. T. Johnsson, V. Negnevitsky,
G. R. Dennis, R. P. Anderson, J. E. Debs, S. S. Szigeti,
K. S. Hardman, S. Bennetts, G. D. McDonald, L. D. Turner,
J. D. Close and N. P. Robins,   New J. Phys. {\bf 15}, 023009 (2013).

\end{thebibliography}
\end{document}